\documentclass{article}
\usepackage{graphicx}

\begin{document}

\title{A class of GUP solutions in deformed quantum mechanics}
\author{Pouria Pedram\thanks{pouria.pedram@gmail.com}\\
{\small Plasma Physics Research Center, Science and Research Campus, Islamic Azad University, Tehran, Iran}}

\date{\today}
\maketitle \baselineskip 24pt

\begin{abstract}
Various candidates of quantum gravity such as string theory, loop
quantum gravity and black hole physics all predict the existence of
a minimum observable length which modifies the Heisenberg
uncertainty principle to so-called Generalized Uncertainty Principle
(GUP). This approach results in the modification of the commutation
relations and changes all Hamiltonians in quantum mechanics. In this
paper, we present a class of physically acceptable solutions for a
general commutation relation without directly solving the
corresponding generalized Schr\"odinger equations. These solutions
satisfy the boundary conditions and exhibit the effect of the
deformed algebra on the energy spectrum. We show that, this
procedure prevents us from doing equivalent but lengthy
calculations.
\end{abstract}


\textit{Pacs}: {04.60.-m}

\section{Introduction}
The quantization of the gravity is one of the oldest problems in the
theoretical physics which has not found a satisfactory solution till
now. Canonical quantization \cite{dewitt} and path integral
quantization \cite{hawking} of gravity are two well-known but old
approaches which try to make the gravity consistent with the laws of
the quantum mechanics. However, the discreteness of the spacetime
which occurs beyond the Planck energy scale is not addressed in
these theories. In fact, the existence of a minimum measurable
length is one of the common features of more recently quantum
gravity theories such as string theory and loop quantum gravity.
Also, we can realize a minimum observable length in the context of
the non-commutativity of the spacetime and some Gedanken experiments
in black hole physics. These studies show that a minimal length of
the order of the Planck length $\ell_{Pl}\approx 10^{-33}cm$ arises
naturally from any theory of the quantum gravity.

Note that, the usual form of the Heisenberg Uncertainty Principle
(HUP) does not predict the existence of a minimum measurable length.
But, HUP breaks down for energies close to the Planck scale where
the corresponding Schwarzschild radius and the Compton wavelength
become approximately equal to the Planck length. Therefore, to
incorporate this idea into quantum mechanics, we need to modify the
uncertainty principle to so-called Generalized Uncertainty Principle
(GUP). This generalization leads us to modify the commutation
relations between the position and the momentum operators in the
Hilbert space. In the last decade, many papers have been appeared in
the literature to address the effects of GUP on the quantum
mechanical systems especially in high energy regime \cite{1}.
Moreover, some authors have recently proposed a GUP which also
implies the existence of a maximum measurable momentum
\cite{main,pedram}. This form of GUP is also consistent with the
doubly special relativity theories \cite{2,3}.

Since in GUP scenario, one cannot probe distances smaller than the
minimum measurable length (string length in the context of string
theory), we expect that it modifies the Hamiltonian of the system
(see \cite{6} and references therein). The modified Hamiltonian
usually contains momentum polynomials as the corrected terms which
in the quantum domain results in the generalized Schr\"odinger
equation. So, the resulting differential equation has completely
different differential structure with respect to the ordinary form
of the Schr\"odinger equation. This makes the problem more complex
especially in the presence of the higher order momentum terms. When
the order of the generalized Schr\"odinger equation increases, we
encounter with many mathematically possible solutions. However,
imposing the physical boundary conditions, reduces the
number of the acceptable solutions.

In this paper, we find the eigenfunctions and eigenvalues of a
particle in a box and a free particle in the context of a
generalized commutation relation in the form $[x,p]=i\hbar(1+\beta
p^{2j})$ where $\beta$ is the GUP parameter and $j=1,2,3,\ldots$. We
demand that these eigenfunctions also satisfy the usual
Schr\"odinger equation but with different eigenvalues. We show that
this condition eliminates unphysical solutions and gives the
modified spectrum without directly solving the differential
equations. Thus, this procedure presents an equivalent but simpler
way to find the solutions of the studied problems.

\section{A Generalized Uncertainty Principle}
Let us start with a generalized uncertainty principle which results
in a minimum observable length
\begin{eqnarray}\label{gup}
 \Delta x \Delta p \geq \frac{\hbar}{2}
\left( 1 +\beta (\Delta p)^2 +\gamma \right),
\end{eqnarray}
where $\beta$ and $\gamma$ are positive quantities which depend on
the expectation value of the position and the momentum operators. In
fact, we have $\beta=\beta_0/(M_{Pl} c)^2$ where $M_{Pl}$ is the
Planck mass and $\beta_0$ is of order the unity. We expect that
these quantities are only relevant in the domain of the Planck
energy $M_{Pl} c^2\sim 10^{19}GeV$. Therefore, in the low energy
regime, we recover the well-known Heisenberg uncertainty principle.
In this limit the parameters $\beta$ and $\gamma$ are irrelevant.
These parameters, in principle, can be obtained from the underlying
quantum gravity theory such as string theory. Note that, the above
relation (\ref{gup}) implies the existence of a minimum observable
length which is equal to $(\Delta x)_{min}=\hbar\sqrt{\beta}$. Since
in the context of the string theory the minimum observable distance
is the string length, we conclude that $\sqrt{\beta}$ is
proportional to this length. In one dimension, Eq.~(\ref{gup})
corresponds to the following commutation relation
\begin{eqnarray}\label{gupc}
[x,p]=i\hbar(1+\beta p^2),
\end{eqnarray}
where the limiting case $\beta=0$ corresponds to the ordinary
quantum mechanics. Note that $x$ and $p$ are symmetric operators on
the dense domain $S_{\infty}$ with respect to the following scalar
product \cite{Kempf}
\begin{eqnarray}
\langle\psi|\phi\rangle=\int_{-\infty}^{+\infty}\frac{dp}{1+\beta
p^2}\psi^{*}(p)\phi(p).
\end{eqnarray}
Moreover, the comparison between Eqs.~(\ref{gup}) and (\ref{gupc})
shows that $\gamma=\beta\langle p\rangle^2$. Now, let us define
\begin{eqnarray}\label{x0p0}
\left\{
\begin{array}{ll}
x = x_{0},\\\\ p = p_{0} \left( 1 + \frac{1}{3}\beta\, p_0^2
\right),
\end{array}
\right.
\end{eqnarray}
where $x_{0}$ and $p_{0}$ obey the canonical commutation relations
$[x_{0},p_{0}]=i\hbar$. One can check that using Eq. (\ref{x0p0}),
Eq. (\ref{gupc}) is satisfied to ${\cal{O}}(\beta)$. Also, from
the above equation we can interpret $p_{0}$ as the momentum operator at
low energies ($p_{0}=-i\hbar
\partial/\partial{x_{0}}$) and $p$ as the momentum operator at
high energies. Now, consider the following form of the Hamiltonian:
\begin{eqnarray}
H=\frac{p^2}{2m} + V(x),
\end{eqnarray}
which using Eq.~(\ref{x0p0}) can be written as
\begin{eqnarray}
H=H_0+\beta H_1+{\cal{O}}(\beta^2),
\end{eqnarray}
where $H_0=\frac{\displaystyle p_0^2}{\displaystyle2m} + V(x)$ and
$H_1=\frac{\displaystyle p_0^4}{\displaystyle3m}$.

To proceed further, let us consider a generalized form of the above
commutation relation:
\begin{eqnarray}\label{gupc2}
[x,p]=i\hbar(1+\beta p^{2j}), \hspace{2cm}j=1,2,3,...,
\end{eqnarray}
where $j=1$ corresponds to our GUP model (\ref{gupc}). For this
general case, we have $p = p_{0} \left( 1 + \frac{\beta}{2j+1}
p_0^{2j} \right)$ and $H_1=\frac{\displaystyle
p_0^{2j+2}}{\displaystyle (2j+1)m}$. Here, we are interested to
present the quantum mechanical solutions of the general form of the
Hamiltonian ($j\geq1$). In the quantum domain, this Hamiltonian
results in the following generalized Schr\"odinger equation in the
quasi-position representation
\begin{eqnarray}\label{H}
-\frac{\hbar^2}{2m}\frac{\partial^2\psi(x)}{\partial
x^2}+\beta\frac{(-1)^{j+1}\hbar^{2j+2}}{(2j+1)m}\frac{\partial^{2j+2}\psi(x)}{\partial
x^{2j+2}} +V(x)\psi(x)=E'\psi(x),
\end{eqnarray}
where the second term is due to the generalized commutation relation
(\ref{gupc2}). This equation is a $2(j+1)$th-order differential
equation which in principle admits $2(j+1)$ independent solutions.
Therefore, solving this equation even for small values of $j$ is
not an easy task. On the other hand, imposing appropriate boundary
conditions, reduces the number of the independent solutions of
Eq.~(\ref{H}). In the next section, for the case of a particle
in a box, we show that the physical solutions also satisfy the
Schr\"odinger equation ($\beta=0$) with different corresponding
eigenenergies. We show that this condition enables us to solve the
generalized Schr\"odinger equation (\ref{H}) without directly
solving the underlying differential equation.

\section{GUP and a particle in a box}
\begin{figure}
\centering
\includegraphics[width=8cm]{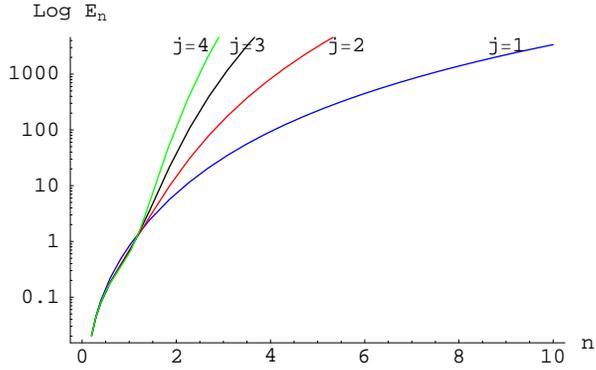}
\caption{Energy spectrum of a particle in a box in the framework of the
generalized commutation relation (\ref{gupc2}) for the first four values of
$j$, $\beta=1$, $m=1$, and $\pi^2\hbar^2/L^2=1$.} \label{fig1}
\end{figure}

Let us consider a particle with mass $m$ confined in an infinite
one-dimensional box with length $L$
\begin{eqnarray}\label{pot}
V(x)=\left\{
\begin{array}{ll}
0 \hspace{1cm} \,\,0<x<L,\\\\ \infty \hspace{1cm}\mbox{elsewhere} .
\end{array}
\right.
\end{eqnarray}
So, the eigenfunctions of a particle in a box should satisfy the
following generalized Schr\"odinger equation
\begin{eqnarray}\label{H2}
-\frac{\hbar^2}{2m}\frac{\partial^2\psi_n(x)}{\partial
x^2}+\beta\frac{(-1)^{j+1}\hbar^{2j+2}}{(2j+1)m}\frac{\partial^{2j+2}\psi_n(x)}{\partial
x^{2j+2}}=E'_n\psi_n(x),
\end{eqnarray}
for $0<x<L$ and they also meet the boundary conditions
$\psi_n(0)=\psi_n(L)=0$. In Ref.~\cite{nozari}, for the case $j=1$,
the above equation is thoroughly solved and its exact eigenvalues and
eigenfunctions are found. Because of the boundary conditions, the
eigenfunctions did not change with respect to the absence of GUP
($\beta=0$). However, the solutions exhibit the effect of GUP on the
eigenvalues which is linear in GUP parameter $\beta$ \cite{nozari}.
These facts lead us to consider the following additional condition
for the eigenfunctions
\begin{eqnarray}\label{H3}
-\frac{\hbar^2}{2m}\frac{\partial^2\psi_n(x)}{
\partial x^2}=E_n\psi_n(x),\hspace{2cm}0<x<L,
\end{eqnarray}
where $E_n=\frac{\displaystyle n^2\pi^2\hbar^2}{\displaystyle
2mL^2}$. If this condition is also satisfied, we can write the
second term in Eq.~(\ref{H2}) in terms of $\psi_n(x)$ \textit{i.e.}
\begin{eqnarray}
\nonumber\frac{\partial^{2j+2} \psi_n(x)}{\partial
x^{2j+2}}&=&\frac{-2mE_n}{\hbar^{2}}\frac{\partial^{2j}\psi_n(x)}{\partial
x^{2j}}=\left(\frac{-2mE_n}{\hbar^{2}}\right)^2\frac{\partial^{2j-2}\psi_n(x)}{\partial
x^{2j-2}}\\
&=&\ldots\,\,\,=\left(\frac{-2mE_n}{\hbar^{2}}\right)^{j+1}\psi_n(x).
\end{eqnarray}
So we have
\begin{eqnarray}\label{H4}
-\frac{\hbar^2}{2m}\frac{\partial^2\psi_n(x)}{\partial
x^2}+\beta\frac{(-1)^{j+1}\hbar^{2j+2}}{(2j+1)m}\frac{\partial^{2j+2}\psi_n(x)}{\partial
x^{2j+2}}=
\left(E_n+\beta\frac{2^{j+1}m^{j}}{(2j+1)}E_n^{j+1}\right)\psi_n(x).
\end{eqnarray}
Now, comparison between Eqs.~(\ref{H2}) and (\ref{H4}) shows that
\begin{eqnarray}
E'_n=E_n+\beta\frac{2^{j+1}m^{j}}{2j+1}E_n^{j+1}.
\end{eqnarray}
For $j=1$ we have $E'_n=\frac{\displaystyle
n^2\pi^2\hbar^2}{\displaystyle
2mL^2}+\frac{\displaystyle\beta}{\displaystyle3}\frac{\displaystyle
n^4\pi^4\hbar^4}{\displaystyle mL^4}$ which is in agreement with the
result of Ref.~\cite{nozari}. Moreover, because of Eq.~(\ref{H3})
the normalized eigenfunctions are
$\psi_n(x)=\sqrt{\frac{\displaystyle 2}{\displaystyle
L}}\sin\left(\frac{\displaystyle n\pi x}{\displaystyle L}\right)$
\cite{nozari}. These results show that in GUP scenario, up to the
first order in $\beta$, there is no change in the eigenfunction but
there is a positive shift in the energy spectrum which is
proportional to $\beta$, $\Delta E_n=\frac{\displaystyle
\beta}{\displaystyle(2j+1)m}\left(\frac{\displaystyle
n^2\pi^2\hbar^2}{\displaystyle L^2}\right)^{j+1}$. Figure \ref{fig1}
shows the effect of GUP on a particle in a box energy levels for the
first four values of $j$.

\section{GUP and a free particle}
In ordinary quantum mechanics the free particle wave function
$u_p(x)$ is defined as the eigenfunction of the momentum operator
$P_{op}$
\begin{eqnarray}\label{p}
P_{op}u_p(x)=p\,u_p(x),
\end{eqnarray}
where $p$ is its eigenvalue. The momentum operator has the following
representation in the position space
\begin{eqnarray}
P_{op}=\frac{\hbar}{i}\frac{\partial}{\partial x}.
\end{eqnarray}
So, from Eq.~(\ref{p}) we have
\begin{eqnarray}\label{p1}
\frac{\hbar}{i}\frac{\partial u_p(x)}{\partial x}=pu_p(x),
\end{eqnarray}
which has the following solution
\begin{eqnarray}\label{solp}
u_p(x)=\frac{1}{\sqrt{2\pi\hbar}}\exp\left({\frac{ip
x}{\hbar}}\right),
\end{eqnarray}
where the constant of integration is chosen to satisfy
$\int^{\infty}_{-\infty}u^{\ast}_p(x)u_p(x')dp=\delta(x-x')$.

In the context of the generalized commutation relation
(\ref{gupc2}), the momentum operator takes the following form in the
position space
\begin{eqnarray}
P_{op}=\frac{\hbar}{i}\frac{\partial}{\partial
x}\left[1+\frac{\beta}{2j+1}\left(\frac{\hbar}{i}\frac{\partial}{\partial
x}\right)^{2j}\right],
\end{eqnarray}
which results in the following eigenvalue equation:
\begin{eqnarray}\label{p2}
\hbar\frac{\partial u_p(x)}{\partial
x}+\beta(-1)^j\frac{\hbar^{2j+1}}{2j+1}\frac{\partial^{2j+1}
u_p(x)}{\partial x^{2j+1}}-ipu_p(x)=0.
\end{eqnarray}
This equation, in principle, has $2j+1$ independent solutions. Now,
consider a class of solutions which satisfy Eqs.~(\ref{p1}) and
(\ref{p2}) at the same time, but with different eigenvalues
($p\rightarrow p'$ in Eq.~(\ref{p1})) \textit{i.e.}
\begin{eqnarray}
u_p(x)=A(p)\exp\left({\frac{ip' x}{\hbar}}\right),
\end{eqnarray}
where $p'=f(p)$. Inserting this solution in Eq.~(\ref{p2}) results
in $p'+\frac{\beta}{2j+1}p'^{2j+1}=p$ which up to the first order in
$\beta$, has the following solution
\begin{eqnarray}
p'=p-\frac{\beta}{2j+1}p^{2j+1},
\end{eqnarray}
or $u_p(x)=A(p)\exp\left(i(p-\frac{\beta}{2j+1}p^{2j+1})
x/\hbar\right)$. To obtain $A(p)$ we demand that the wave function
satisfies the normalization condition
$\int^{\infty}_{-\infty}u^{\ast}_p(x)u_p(x')dp=\delta(x-x')$ which
results in $A(p)=\left(\frac{1-\beta
p^{2j}}{2\pi\hbar}\right)^{1/2}$. So, the eigenfunctions of the
generalized momentum operator, up to the first order in $\beta$,
take the following form
\begin{eqnarray}
u_p(x)=\left(\frac{1-\beta
p^{2j}}{2\pi\hbar}\right)^{1/2}\exp\left(\frac{i(p-\frac{\beta}{2j+1}p^{2j+1})x}{\hbar}\right).
\end{eqnarray}
This result coincides with the previous solution for the case $j=1$
which was found after ``lengthy'' calculations \cite{nozari}. Note
that this solution for $\beta\rightarrow0$ reduces to (\ref{solp})
in order to satisfy the correspondence principle.

\section{Conclusions}
We have studied the effects of GUP on the spectrum of two quantum
mechanical systems. We found a class of GUP solutions for a particle
in a box and a free particle in the context of a general form of
deformed position and momentum commutation relations. To obtain the
physically acceptable solutions, we demanded that the sought-after
eigenfunctions satisfy both the ordinary and generalized
Schr\"odinger equations at the same time but with different
corresponding eigenvalues. We showed that this method is equivalent
with directly solving the generalized Schr\"odinger equation after imposing
the physically acceptable boundary conditions. In fact, this
procedure prevents us from doing equivalent but lengthy
calculations.

\end{document}